\documentclass[aps,pra,showpacs,notitlepage]{revtex4-1}

\usepackage{graphicx}
\begin{document}

\title{The role of system-meter entanglement in controlling the resolution\\ and decoherence of quantum measurements}

\author{Kartik Patekar}
\affiliation{
Department of Physics, Indian Institute of Technology Bombay, Mumbai 400076, India
}
\author{Holger F. Hofmann}
\email{hofmann@hiroshima-u.ac.jp}
\affiliation{
Graduate School of Advanced Sciences of Matter, Hiroshima University,
Kagamiyama 1-3-1, Higashi Hiroshima 739-8530, Japan
}

\begin{abstract}
Measurement processes can be separated into an entangling interaction between the system and a meter and a subsequent readout of the meter state that does not involve any further interactions with the system. In the interval between these two stages, the system and the meter are in an entangled state that encodes all possible effects of the readout in the form of non-local quantum correlations between the system and the meter. 
Here, we show that the entanglement generated in the system-meter interaction expresses a fundamental relation between the amount of decoherence and the conditional probabilities that describe the resolution of the measurement. Specifically, the entanglement generated by the measurement interaction correlates both the target observable and the back-action effects on the system with sets of non-commuting physical properties in the meter. The choice of readout in the meter determines the trade-off between irreversible decoherence and measurement information by steering the system into a corresponding set of conditional output states. The Hilbert space algebra of entanglement ensures that the irreversible part of the decoherence is exactly equal to the Hellinger distance describing the resolution achieved in the measurement. We can thus demonstrate that the trade-off between measurement resolution and back-action is a fundamental property of the entanglement generated in measurement interactions.
\end{abstract}

\maketitle
%%--Introduction

\section{Introduction}

One of the fundamental conceptual problems in quantum mechanics is the role of measurements in the definition of physical reality. Originally, it was simply assumed that the measurement process could be described by classical arguments, so that the Hilbert space formalism can be justified in terms of established assumptions about physical objects in space and time. This historical line of argument is  particularly obvious in Heisenberg's initial discussion of the uncertainty principle, which avoids a detailed discussion of the system-meter interaction and focuses only on the Hilbert space of the system \cite{Hei27}. Although it was soon recognized that a consistent formulation of the theory requires a complete quantum mechanical description of the measurement process, the flaws of Heisenberg's semiclassical approach to quantum measurements were not really addressed until quantum optics provided the experimental means of realizing a fully quantum mechanical which-path measurement inside an interferometric set up, resulting in a rather heated controversy over the relation between uncertainty and complementarity \cite{Scu89,Scu91,Sto94,Eng95,Wis95,Dur98}. A widespread impression at the time was that textbook definitions of measurement uncertainties were inadequate. This impression was strengthened by the state-dependent analysis of measurement uncertainties introduced by Ozawa \cite{Oza03}, resulting in yet another round of criticisms and controversy \cite{Wat11,Bus13,Dre14,Bus14,Roz15,Mao19}. At the same time, measurement theories attained new relevance in the context of quantum information, where the focus shifted from uncertainties towards quantum state discrimination \cite{Fuc96,Mac06,Bcm14,Ari19}. As a result of all of these developments, there is now an abundance of methods and approaches to quantum measurement that has made it even more difficult to find any common ground on fundamental questions regarding the role and the significance of the measurement process in quantum theory. 
%%--added comment (change 1)
This confusion is all the more regrettable since it is becoming increasingly clear that joint and sequential measurements are an important tool in the study of non-classical correlations \cite{Gog11,Suz12,Opp10,Ban13,Lod17,Hof19}, indicating that a better understanding of the measurement process might shed some light on the essential resources used in quantum information technologies.
%%--
It would therefore be good to trace the problem back to its origin and ask why Heisenberg's semiclassical approach to measurement uncertainties failed. In fact, the answer to this question was already implied in the earliest paper on complementarity in quantum measurements \cite{Scu89}. The reason why a semiclassical approach to quantum measurements must fail is that ideal measurements require an entangling system-meter interaction as a first step in a two step process, invalidating any attempt to describe the meter as part of a macroscopic and hence classical environment. We therefore believe that the key to a proper understanding of quantum measurements is a clear and unambiguous separation between this first step and the subsequent second step described by the readout process. This separation allows us to identify the statistical limits of quantum measurements with the non-classical statistical limits of the entanglement generated in the first step of the measurement process, where the second step describes how the non-classical correlations generated by the entangling interaction condition the output state of the system.

In the present paper, we formulate a general description for quantum measurements of a specific target observable $\hat{A}$ performed on an arbitrary input state of the system. We point out that the entangling interaction itself disturbs the state of the system by converting local quantum coherences of the system into non-local correlations between the system and the meter. Importantly, correlations between the target observable and the meter system coexist with complementary correlations between the disturbance of coherences in the system and a corresponding meter observable. Instead of deciding the resolution-disturbance trade-off, the overall disturbance of coherence caused by the entangling interaction merely defines an upper limit for the resolution that could be achieved when the meter is read out. An analysis of the entanglement shows that there is a precise mathematical relation between the ability to distinguish eigenstates of $\hat{A}$ and the decoherence between these eigenstates when the resolution is quantified in terms of the Hellinger distance between the conditional probabilities of the measurement outcomes. The Hellinger distance therefore provides a natural measure of resolution for quantum measurements of an observable $\hat{A}$. However, the entangling interaction only determines the upper limit of the Hellinger distance between the conditional probability distributions. The actual resolution is determined by the readout strategy, which selects a specific balance between the meter properties correlated with the eigenstates of $\hat{A}$ and the meter properties correlated with the coherent phase changes in the system that describe the disturbance effects of the measurement. Essentially, the entanglement generated in the measurement interaction makes it possible to steer the quantum measurement between a maximal resolution achieved when no information about the phase changes is obtained and the disturbance is completely irreversible and a possible erasure of all measurement information that recovers the full coherence of the input state, making the disturbance completely reversible \cite{Scu82}. The fundamental properties of the entanglement thus fully determine the trade-off between measurement resolution and irreversible disturbance in the readout step of the measurement process. 

Our analysis shows that the problem of measurement uncertainty is fundamentally the same as the problem of quantum steering using entangled states. The disturbance of the initial system state is a necessary characteristic of entanglement generation in a fully reversible unitary interaction between the system and the meter. The theory presented in this paper shows how the information transfer described by the system-meter interaction can be characterized with a minimum of assumptions about the properties of the meter system itself. It is then possible to clearly distinguish between the role of the entangling interaction and the different readout strategies realized in the meter system only. We can thus show that, as a result of the presence of entanglement, the actual selection of the post-measurement quantum state is completely independent of the physics of the system. From these observations, we conclude that the role of entanglement in quantum measurement is the decoupling of the selection of output states from the physics of the system, indicating that there is no context-independent description of the emergence of specific output states in quantum measurements.

The rest of the paper is organized as follows.
In Sec. \ref{Sec:inter}, we introduce a compact formulation for the minimal physical interaction between the system and the meter required in a measurement of a specific observable $\hat{A}$. It is shown that the quantum mechanical part of the measurement represented by the measurement interaction can be characterized completely by a set of conditional states $\{\mid \phi(a) \rangle_M\}$. In Sec. \ref{Sec:resolve}, we discuss the concept of measurement resolution and show that it can be described by the Hellinger distance between conditional probabilities for different eigenstates of $\hat{A}$. The upper bound for the Hellinger distances between $a_1$ and $a_2$ is determined by the quantum state overlap of the conditional states $\mid \phi(a_1) \rangle$ and $\mid \phi(a_2) \rangle$. In Sec. \ref{Sec:disturb}, we analyze the decoherence caused by the entangling interaction and show that the relative reduction of the off-diagonal elements of the density matrix is equal to the upper limit of the resolution defined by the Hellinger distance. In Sec. \ref{Sec:readout}, we consider the resolution and the conditional output states associated with a specific readout strategy and show that the irreversible part of the decoherence is precisely equal to the resolution for maximally coherent interactions and readouts. In Sec. \ref{Sec:steering}, it is shown that the selection of a readout strategy steers the conditional output state of the system between high resolution and high coherence in such a way that the presence of entanglement can be verified by the violation of a steering inequality. In Sec. \ref{Sec:character}, we consider possible criteria for the construction of optimal measurement interactions. It is shown that a resolution at the limit set by the total decoherence can be obtained if and only if there exists a representation of the conditional states $\{\mid \phi(a) \rangle_M\}$ with real and positive inner products. Sec. \ref{Sec:conclude} concludes the paper. 

\section{Measurement interactions for a specific target observable}
\label{Sec:inter}

In the original formulation of quantum theory, it is assumed that the eigenstates $\mid a \rangle$ of an operator observable $\hat{A}$ represent the different measurement outcomes obtained in a precise measurement of $\hat{A}$. On closer inspection, this assumption is justified because all measurements require some form of interaction to produce an observable effect of the property $\hat{A}$ of the system on the apparatus used as a meter. In the following, we consider an interaction between the system $S$ and the meter $M$ that is ideally suited for a measurement of the observable $\hat{A}$. The essential property of such an interaction is that it should not perturb the eigenstates of the property $\hat{A}$ in the system. We can then represent the initial system-meter interaction by a unitary transformation $\hat{U}_{SM}$, where the requirement that the eigenstates $\mid a \rangle$ are unperturbed results in the definition of conditional unitaries,
\begin{equation}
\label{eq:condU}
\hat{U}_{SM} \mid a \rangle_S = \mid a \rangle_S \otimes \hat{U}_M(a).
\end{equation}
Eq.(\ref{eq:condU}) shows that it is possible to completely characterize the measurement interaction in terms of the conditional unitaries $\hat{U}_M(a)$ operating on the meter system. In addition, we know that the meter will be initialized in a specific state $\mid \Phi_0 \rangle_M$ before each measurement. Since we are not interested in the details of the meter, we can summarize the initial state and the effects of the conditional unitaries to obtain the effect of the interaction in terms of the output state components,
\begin{equation}
\label{eq:condstates}
\hat{U}_{SM} \mid a \rangle_S \mid \Phi_0 \rangle_R = \mid a \rangle_S \mid \phi(a) \rangle,
\end{equation}
where the conditional meter states $\mid \phi(a) \rangle$ are found by applying the conditional unitaries $\hat{U}_M(a)$ to the initial meter state $\mid \Phi_0 \rangle_M$. 

\begin{figure}[th]
\vspace{-2cm}
\begin{picture}(500,360)
\put(50,0){\makebox(360,360){
\scalebox{0.6}[0.6]{
\includegraphics{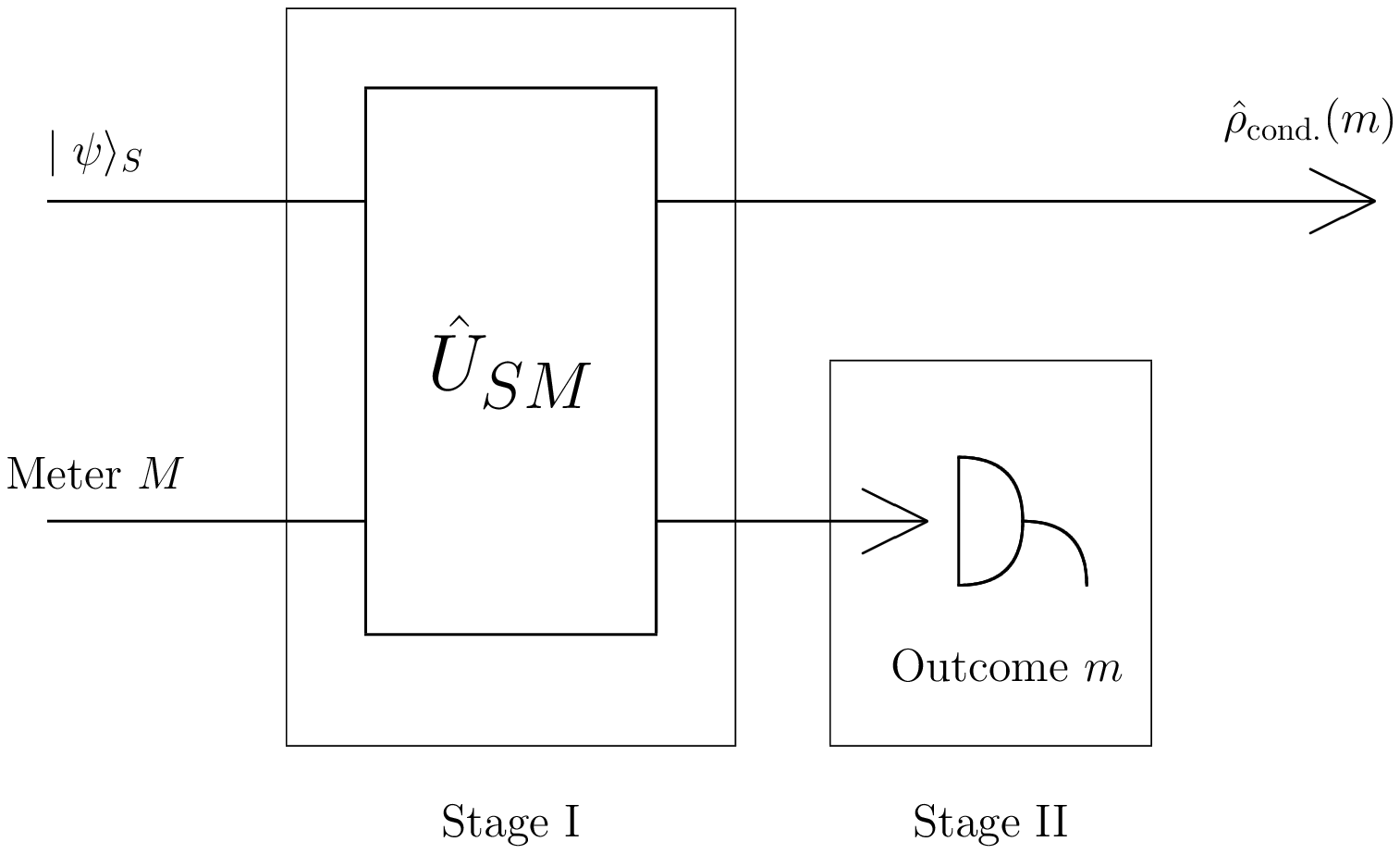}}}}
\end{picture}
\vspace{-2.5cm}
\caption{\label{fig1}
Schematic representation of a quantum measurement as a sequence of entangling system-meter interaction (stage I) and meter readout (stage II). The interaction $\hat{U}_{SM}$ determines the quantum correlations between the system and the meter that make system information available at the readout state while simultaneously changing the quantum coherence of the input state. 
}
\end{figure}

By describing the initial measurement interaction in terms of a set of conditional meter states $\mid \phi(a) \rangle_M$, we can capture the essential process of information transfer from the system to the meter in its most compact quantum mechanical form. However, the objective description of a quantum measurement is not complete until the information extracted from the system is converted into an irreversibly recorded output signal. The merit of our analysis is that we can describe the actual physical interaction between the system and the meter in maximally coherent quantum mechanical terms, so that the readout stage of the measurement can be confined to the meter. Fig. \ref{fig1} illustrates this decomposition of the measurement process into interaction and readout. In general, the readout stage can be represented by a measurement basis $\mid m \rangle_M$ in the Hilbert space of the meter, where the probabilities for a measurement result of $m$ conditioned by the state $\mid \phi(a) \rangle_M$ are given by $P(m|a)=|\langle m \mid \phi(a) \rangle|^2$. It is therefore possible to discuss the effects of different readout strategies in terms of the Hilbert space geometry defined by the conditional states $\mid \phi(a) \rangle_M$ of the meter system that originated from the interaction stage of the measurement process. In the following, we will take a closer look at the relation between the measurement information obtained in the readout and the Hilbert space geometry of the conditional meter states associated with different eigenstates of the target observable $\hat{A}$.

\section{Measurement resolution}
\label{Sec:resolve}

%%--change 2: addition of references
Modern measurement theory makes it difficult to decide on a proper definition of resolution. In general, information theoretic approaches tend to focus on quantum state discrimination \cite{Hei27,Oza03}, while uncertainty based approaches focus on quantitative fluctuations of the operator observable \cite{Fuc96}. For our purposes, it is necessary to select a detailed characterization of the relation between the input statistics $P(a)=|\langle a \mid \psi \rangle|^2$ in the system and the output statistics $P(m)$ in the meter. We therefore define the resolution as a function of the conditional probability distributions $P(m|a)$, similar to the approach proposed by Buscemi et al. as an information theoretic approach to noise and disturbance in \cite{Bcm14}. However, we will try to retain a more detailed microscopic description by defining the resolution as the ability to distinguish between a specific pair of eigenstates in the input. The resolution will then be a symmetric matrix of the eigenstates $\mid a \rangle$ given by its elements $R(a_1,a_2)$. As pointed out above, the conditional probabilities that need to be resolved are given by 
\begin{equation}
\label{eq:condprob}
P(m|a)=|\langle m \mid \phi(a) \rangle|^2.
\end{equation}
This means that the resolution $R(a_1,a_2)$ must be a measure of the statistical distance between the two conditional probability distributions $P(m|a_1)$ and $P(m|a_2)$. Since we are interested in a measure that connects well with quantum mechanical features of Hilbert space inner products, we choose the squared Hellinger distance, defined as
\begin{equation}
R(a_1,a_2) = \frac{1}{2} \sum_m \left(\sqrt{P(m|a_1)}-\sqrt{P(m|a_2)}\right)^2.
\end{equation}
Although the Hellinger distance is a classical statistical distance, its use of the square roots of probabilities invites a comparison with Hilbert space vectors and their inner products. Specifically, the inner product of the vectors $\sqrt{P(m|a_1)}$ and $\sqrt{P(m|a_2)}$ is known as the Bhattacharyya coefficient and corresponds to a real valued version of the Hilbert space inner product. In the present case, the Bhattacharyya coefficient is related to the conditional meter states by
\begin{equation}
\sum_m \sqrt{P(m|a_1) P(m|a_2)} = \sum_m |\langle \phi(a_1) \mid m \rangle \langle m \mid \phi(a_2) \rangle|. 
\end{equation}
The only difference between the right hand side of this equation and the inner product is the use of absolute values before the summation over $m$. Since any variation of phases in the sum can only diminish the total result, the inner product of the conditional meter states is a lower bound of the Bhattacharyya coefficient for all possible readout measurements $\{ \mid m \rangle\}$,
\begin{equation}
\sum_m \sqrt{P(m|a_1) P(m|a_2)} \geq |\langle \phi(a_1) \mid \phi(a_2) \rangle|. 
\end{equation}
For the squared Hellinger distance, it follows that 
\begin{equation}
\label{eq:Rmatrix}
R(a_1,a_2) \leq 1-|\langle \phi(a_1) \mid \phi(a_2) \rangle|.
\end{equation}
The squared Hellinger distance therefore relates the resolution between $a_1$ and $a_2$ to the absolute value of the quantum state overlap between the conditional meter states $\mid \phi(a_1) \rangle$ and $\mid \phi(a_2) \rangle$. In this manner, it is possible to trace the measurement resolution back to the initial interaction. If the interaction was weak, the conditional states are still nearly equal and have inner products close to one. High resolution requires an interaction that results in lower values of the inner products. 

Strictly speaking, the measurement resolution $R(a_1,a_2)$ is determined by comparing the measurement statistics for the input eigenstates $\mid a_1 \rangle$ and $\mid a_2 \rangle$. However, the same measurement process can also be applied to arbitrary superpositions of the eigenstates $\mid a \rangle$. In that case, the measurement interaction results in an entangled state of the system and the meter, with non-classical correlations that are described entirely in terms of the conditional quantum state components $\mid \phi(a) \rangle$. As we shall show in the following, this generation of entanglement by the measurement interaction represents the physics of disturbance in a quantum measurement. 

\section{Entanglement and disturbance}
\label{Sec:disturb}

The original formulation of measurement uncertainties put forward by Heisenberg in \cite{Hei27} was based on a semiclassical model of the interaction and neglected the role of entanglement in any fully quantum mechanical description of the interaction. This is the reason why it has been rather difficult to formalize the concept of disturbance in quantum measurements in a commonly accepted manner \cite{Rod19}. Here, we will propose a more direct characterization of disturbance in terms of the actual changes of the quantum state described by the entangling interaction given in Eq.(\ref{eq:condstates}) above. 
Since the eigenstates of the target observable $\hat{A}$ are also eigenstates of the unitary interaction $\hat{U}_{SM}$, the disturbance caused by the interaction dynamics must necessarily appear in the coherences between different eigenstates of $\hat{A}$. We therefore need to consider the effect of the unitary interaction $\hat{U}_{SM}$ on an arbitrary superposition $\mid \psi \rangle$ of the eigenstates $\mid a \rangle$,
\begin{equation}
\label{eq:entangle}
\hat{U}_{SM} \mid \psi \rangle_S \mid \Phi_0 \rangle_R = \sum_a \langle a \mid \psi \rangle \mid a \rangle_S \mid \phi(a) \rangle_R.
\end{equation}
It should be obvious that the output state describes entanglement between the system and the meter. Importantly, this entanglement is an unavoidable consequence of the application of a measurement interaction to a superposition of target observable eigenstates. In this sense, entanglement generation is a more fundamental feature of quantum measurements than disturbance. Indeed, the disturbance of the initial state is merely a necessary byproduct of entanglement generation. If we trace out the meter system, the output state of the system is described by a density operator $\hat{\rho}_S(\mbox{out})$, where the density matrix elements are related to their input values by
\begin{equation}
\label{eq:decohere}
\langle a_1 \mid \hat{\rho}_S(\mbox{out}) \mid a_2 \rangle =
\langle \phi(a_2) \mid \phi(a_1) \rangle \langle a_1 \mid \psi \rangle \langle \psi \mid a_2 \rangle.
\end{equation}
The unconditional disturbance of the system state by the entangling measurement interaction therefore consists of a reduction of phase coherences by a factor given by the inner product of the conditional meter states associated with the quantum state components $\mid a_1 \rangle$ and $\mid a_2 \rangle$. Independent of the input state, we can quantify the decoherence caused by the measurement interaction in the system using the ratio of the coherences of the input density operator $\hat{\rho}_S(\mbox{in})=\mid \psi \rangle \langle \psi \mid$ and the output density operator $\hat{\rho}_S(\mbox{out})$. The reduction of coherence is then given by the decoherence matrix,
\begin{equation}
D(a_1,a_2) = 1 - \left|\frac{\langle a_1 \mid \hat{\rho}_S(\mbox{out}) \mid a_2 \rangle}{\langle a_1 \mid \hat{\rho}_S(\mbox{in}) \mid a_2 \rangle}\right|.
\end{equation}
According to Eq.(\ref{eq:decohere}), the decoherence caused by the measurement interaction is given by the inner products of the conditional meter states,
\begin{equation}
\label{eq:Dmatrix}
D(a_1,a_2) = 1 - |\langle \phi(a_2) \mid \phi(a_1) \rangle|.
\end{equation}
Comparison with Eq.(\ref{eq:Rmatrix}) shows that these decoherence factors are equal to the upper bound of the squared Hellinger distance describing the maximal resolution between two eigenvalues of the target observable,
\begin{equation}
\label{eq:RDUR}
R(a_1,a_2) \leq D(a_1,a_2).
\end{equation}
This seems to be the most precise formulation of the resolution-disturbance trade-off in quantum measurements, relating the ability to distinguish between eigenstates of $\hat{A}$ by a meter readout to the necessary loss of quantum coherence between these eigenstates caused by the measurement interaction.
%%--change 2, additional explanation
We achieve this higher level of precision by defining a separate resolution for each pair of eigenstates, which is naturally related to the loss of coherence between these eigenstates. It might be worth noting that this approach is fundamentally different from all of the global measures for resolution and disturbance considered in the references \cite{Oza03,Wat11,Bus13,Dre14,Bus14,Roz15,Mao19,Fuc96,Mac06,Bcm14,Ari19} mentioned in the introduction. It is therefore quite possible that the detailed description of resolution and decoherence given here is the essential key to a more fundamental understanding of the uncertainty trade-off in quantum measurements. Specifically, the present approach avoids two problems that have made it difficult to identify more simple relations between resolution and disturbance. The first problem is that the statistics of measurement resolution are quite complicated and any global measure obtained from averages over all possible eigenstates must necessarily omit much of the details that characterize the actual distribution of measurement errors. The second problem is that a definition of disturbance in terms of an additional observable of the system is a rather arbitrary choice that is not directly related to the dynamics of the actual measurement. Much of the discussion surrounding the concept of disturbance in quantum measurements seems to originate from this problem and its consequences \cite{Rod19}.
%%---

For pure state inputs, both the maximal achievable resolution and the associated decoherence are features of an entangled state that describes the quantum correlations between the meter and the system generated in the measurement interaction. This entangled state completely describes the first stage of the measurement process, as shown in Fig. \ref{fig1}. In the second stage of the measurement process, the result will be read out in the meter system only, where the entanglement generated in the first stage describes the statistical correlations between the physical properties in the system output and the meter information actually obtained in the readout. The fact that the meter and the system are entangled means that there is yet another trade-off involved in the selection of the meter readout. This trade-off corresponds to the possibility of steering the quantum coherences of the remote system of an entangled pair by choosing between different measurement strategies in the local system \cite{Wis07}. 

\section{Readout and irreversible decoherence}
\label{Sec:readout}

The presence of entanglement after the interaction indicates that the output state of the system will be conditioned by the readout measurement in the meter system. Specifically, entanglement means that the ``collapse'' of the wavefunction associated with a measurement does not happen within the system, but needs to be described in terms of the quantum correlations established between the system and the meter. Note that this may give the somewhat misleading impression that the disturbance of the input state is determined in some mysterious non-local manner by the physics of the meter. Indeed, this confusion may well have been the reason for the initial controversy over measurement uncertainties \cite{Scu89,Scu91,Sto94,Eng95,Wis95,Dur98}. However, it is highly problematic to compare the conditional output state with the input state, since the change of the probabilities of $a$ between the input state and the conditional output state merely represent a Bayesian update associated with the conditional probabilities $P(m|a)$ \cite{Roz15}. To avoid the possible confusion between Bayesian updates and physical changes of the state, it is necessary to always compare the complete statistics of $m$ with the input state. 

The output state $\hat{\rho}_{\mathrm{cond.}}(m)$ conditioned by a readout result $m$ can be given by its density matrix elements,
\begin{equation}
\langle a_1 \mid \hat{\rho}_{\mathrm{cond.}}(m) \mid a_2 \rangle =
\frac{\langle \phi(a_2) \mid m \rangle \langle m \mid \phi(a_1) \rangle}{p(m)} \langle a_1 \mid \hat{\rho}_S(\mbox{in}) \mid a_2 \rangle,
\end{equation}
where the outcome probabilities $p(m)$ are given by the input probabilities of $a$,
\begin{equation}
p(m) = \sum_a p(m|a) \langle a \mid \hat{\rho}_S(\mbox{in}) \mid a \rangle.
\end{equation}
It is important to note that the readout does not change the outcome independent quantum statistics of the system, as confirmed by a sum over all possible outcomes $m$,
\begin{equation}
\label{eq:reversibleD}
\sum_m p(m) \hat{\rho}_{\mathrm{cond.}}(m) = \hat{\rho}_S(\mbox{out}).
\end{equation}
The readout does not change the total disturbance caused by the measurement interaction, but it can provide information about the precise magnitude of the phase changes in the off-diagonal elements of the density matrix describing the system. This means that the coherence does not simply disappear as the quantum phases are randomized. Instead, each readout $m$ has its own phase, and the phase shifts associated with each readout $m$ can be undone by a conditional unitary transformation performed on the system, as demonstrated in quantum eraser measurements \cite{Scu82,Yoo00}. Due to the entanglement between the system and the meter, information about both the target observable $\hat{A}$ and its coherences in the system are available in the meter degrees of freedom and no irreversible measurement or decoherence has occurred until the readout process is fixed by additional dynamics confined to the meter only. It is therefore important to distinguish between the reversible decoherence defined by the entangling system-meter interaction and represented by the sum in Eq. (\ref{eq:reversibleD}), and the irreversible decoherence defined by a specific choice of readout in the meter and represented by the conditional density matrices of that readout. 

To evaluate the irreversible part of the decoherence $D(a_1,a_2)$ for a specific measurement readout, it is convenient to sum up the absolute values of the off-diagonal elements in the conditional density matrices, which provides an appropriate measure of quantum coherences for statistical averages \cite{Bau14}. Since a local unitary transformation could be used to restore the original input state phases, this is also the minimal decoherence that remains when the readout information is used to compensate the decoherence by unitary feedback on the system, 
\begin{equation}
\label{eq:irreversible}
D_{\mathrm{irr.}}(a_1,a_2) = 1 - \frac{\sum_m p(m) |\langle a_1 \mid \hat{\rho}_{\mathrm{cond.}}(m)\mid a_2 \rangle| }{|\langle a_1 \mid \hat{\rho}_S(\mbox{in}) \mid a_2 \rangle|}.
\end{equation}
The sum of the absolute values of the off-diagonal elements can be expressed in terms of the conditional probabilities $P(m|a)=|\langle m \mid \phi(a)\rangle|^2$,
\begin{equation}
\sum_m p(m) |\langle a_1 \mid \hat{\rho}_{\mathrm{cond.}}(m)\mid a_2 \rangle| = \left(\sum \sqrt{P(m|a_1) P(m|a_2)}\right) \langle a_1 \mid \hat{\rho}_S(\mbox{in}) \mid a_2 \rangle.
\end{equation} 
The irreversible part of the decoherence given by Eq.({\ref{eq:irreversible}) is therefore exactly equal to the squared Hellinger distance describing the measurement resolution achieved by the present readout strategy,
\begin{equation}
\label{eq:tradeoff}
D_{\mathrm{irr.}}(a_1,a_2) = R(a_1,a_2).
\end{equation}
For maximally coherent measurement interactions and readout, the trade-off between resolution and irreversible disturbance of the system is therefore exact. Specifically, the entanglement generated by the measurement interaction guarantees that a quantum mechanically precise readout of the meter either provides information about $a$, or about the change of coherent phases caused by the interaction. This trade-off between measurement information and information about the disturbance of the system caused by the measurement interaction in the readout stage of the measurement characterized the role of system-meter entanglement in the measurement process. We will therefore take a closer look at the actual entanglement generated in the measurement interaction by evaluating the input state dependent entanglement using a steering inequality that includes the state independent concepts of measurement resolution and irreversible decoherence. 

\section{Quantum steering}
\label{Sec:steering}

Different readout strategies applied to the meter system will result in different conditional output states in the system. This is direct evidence of entanglement, corresponding to the idea of quantum steering \cite{Wis07}. Specifically, the presence of entanglement after the interaction between system and meter permits the selection of different sets of mutually incompatible conditional states, depending on the readout measurement performed in the meter system only. To characterize this fundamental role of entanglement in quantum measurements, we need to analyze the conditional quantum statistics of different readout strategies in more detail. 

It is in fact possible to characterize the relation between the conditional output states and the input state entirely in terms of the resolution and the irreversible decoherence of the specific readout used in stage two of the quantum measurement. For the resolution, we can consider the diagonal elements of the conditional output matrices,
\begin{equation}
\langle a \mid \hat{\rho}_{\mathrm{cond.}}(m)\mid a \rangle = \frac{p(m|a)}{p(m)} \langle a \mid \hat{\rho}_S(\mbox{in}) \mid a \rangle.
\end{equation}
The resolution matrix can be obtained from the conditional output matrices by averaging over the square root of the products of two diagonal elements,
\begin{equation}
\label{eq:Rstate}
\sum_m p(m) \sqrt{\langle a_1 \mid \hat{\rho}_{\mathrm{cond.}}(m)\mid a_1 \rangle \langle a_2 \mid \hat{\rho}_{\mathrm{cond.}}(m)\mid a_2 \rangle} = (1-R(a_1,a_2)) \sqrt{\langle a_1 \mid \hat{\rho}_S(\mbox{in}) \mid a_1 \rangle\langle a_2 \mid \hat{\rho}_S(\mbox{in}) \mid a_2 \rangle}.
\end{equation}
Likewise, the decoherence matrix can be obtained from the average coherence of the conditional density matrices,
\begin{equation}
\label{eq:Dstate}
\sum_m p(m) |\langle a_1 \mid \hat{\rho}_{\mathrm{cond.}}(m)\mid a_2 \rangle| = (1-D_{\mathrm{irr.}}(a_1,a_2)) |\langle a_1 \mid \hat{\rho}_S(\mbox{in})  \mid a_2 \rangle|.
\end{equation}
It is therefore possible to determine the resolution and the irreversible decoherence from the relation between the input state and the conditional output states obtained for the different measurement outcomes $m$. Since each of the conditional density matrices must satisfy positivity, the square root products of diagonal elements limit the coherences to
\begin{equation}
\label{eq:positive}
|\langle a_1 \mid \hat{\rho}_{\mathrm{cond.}}(m)\mid a_2 \rangle|
\leq
\sqrt{\langle a_1 \mid \hat{\rho}_{\mathrm{cond.}}(m)\mid a_1 \rangle \langle a_2 \mid \hat{\rho}_{\mathrm{cond.}}(m)\mid a_2 \rangle}.
\end{equation}
%%--change 3, improved explanation of the inequality for conditional state positivity
Since the positivity of the conditional density matrices must be guaranteed for all possible input states, it is possible to relate Eq.(\ref{eq:positive}) to the limits on resolution and decoherence by expressing both sides in terms of the corresponding input state expressions given by Eqs.(\ref{eq:Rstate}) and (\ref{eq:Dstate}),
\begin{equation}
\label{eq:steer1}
(1-D_{\mathrm{irr.}}(a_1,a_2)) |\langle a_1 \mid \hat{\rho}_S(\mbox{in})  \mid a_2 \rangle|
\leq 
(1-R(a_1,a_2)) \sqrt{\langle a_1 \mid \hat{\rho}_S(\mbox{in}) \mid a_1 \rangle\langle a_2 \mid \hat{\rho}_S(\mbox{in}) \mid a_2 \rangle}
.
\end{equation}
In the limit of maximally coherent interactions and readout, Eq.(\ref{eq:tradeoff}) ensures that this inequality is satisfied for all positive input states. The positivity of all conditional density matrices therefore depends on the validity of Eq.(\ref{eq:tradeoff}), or more generally on the necessary requirement that the upper bound of the resolution $R(a_1,a_2)$ is given by the irreversible decoherence $D_{\mathrm{irr.}}(a_1,a_2)$ in accordance with the general uncertainty bound of Eq.(\ref{eq:RDUR}). As Eq.(\ref{eq:tradeoff}) shows, this requirement is automatically satisfied when every single readout strategy is considered separately. 

In the absence of system-meter entanglement, Eq.(\ref{eq:steer1}) would also restrict the relation between the resolutions and the irreversible decoherences of different readout strategies. The role of entanglement in controlling the values of resolution and decoherence can therefore be verified by the violation of a steering inequality based on the possibility of readout strategies where the left hand side of the inequality for one strategy is higher than the right hand side of the inequality for another strategy.
%%----
If the fist stage of the measurement could be described by a classical meter, the readout information would have to be present in the meter regardless of the readout strategy chosen in the meter system. If we consider two readout strategies, $m_r$ and $m_c$, a non-entangling measurement interaction would have to produce conditional output states $\hat{\rho}_{\mathrm{cond.}}(m_r,m_c)$ conditioned by both readout results, and each of these conditional density matrices would have to satisfy Eq.(\ref{eq:positive}). Each readout strategy would therefore represent an incomplete readout, with an unnecessarily low resolution and an unnecessarily high irreversible decoherence,
\begin{eqnarray}
R(a_1,a_2) &\leq& \mbox{Max}(R(a_1,a_2))
\nonumber \\
D_{\mathrm{irr.}}(a_1,a_2) &\geq& \mbox{Min}(D_{\mathrm{irr.}}(a_1,a_2)).
\end{eqnarray}
In the absence of entanglement, it would be impossible to violate Eq.(\ref{eq:steer1}) by using different readout strategies on the left hand side and on the right hand side, since the different conditional density matrices associated with each readout strategy could all be explained by a single set of conditional quantum states. If the left hand side readout $m_c$ minimizes the irreversible decoherence and the right hand side readout $m_r$ maximizes the resolution, a non-entangling measurement interaction therefore requires that the relation between the maximal resolution and the minimal irreversible decoherence is given by
\begin{equation}
\label{eq:steer2}
\mbox{Max}(R(a_1,a_2)) \leq \mbox{Min}(D_{\mathrm{irr.}}(a_1,a_2)).
\end{equation}
Any violation of this inequality would result in a violation of Eq.(\ref{eq:steer1}) for the classically allowed case of a joint readout of $m_c$ and $m_r$. However, quantum measurements necessarily violate this inequality, because both the resolution and the decoherence are complementary aspects of a single entangled state formed in the interaction between the system and the meter. 

%%--change 4, comment on LGI
It should be noted that there is no contradiction between quantum steering and local realism, since the assumption that the conditional density matrices of the system must be positive is based on standard quantum mechanics. This means that a violation of Eq.(\ref{eq:steer2}) can be perfectly consistent with a hidden variable description of measurement errors such as the spin flip model used to characterize the violation of Leggett-Garg inequalities in \cite{Suz12}. On the other hand, the uncertainty limit given by Eq.(\ref{eq:RDUR}) must be satisfied by each individual readout strategy in order to avoid the emergence of negative probabilities in a sequential measurement of the observables violating Leggett-Garg inequalities. The violation of Eq.(\ref{eq:steer2}) thus shows that the local failure of joint measurability described by Leggett-Garg inequalities can be converted into a failure of local realism in the correlation between the system and the meter used for the intermediate measurement in the Leggett-Garg scenario described in \cite{Suz12}. The possibility of steering in quantum measurements may therefore help to shed new light on the fundamental relation between measurement uncertainties and the limits of non-classical statistics in quantum systems \cite{Gog11,Suz12,Opp10,Ban13,Lod17,Hof19}.
%%--

%%--change 5, explicit steering inequality violation
The role of entanglement in quantum measurement can be verified experimentally by confirming that the steering inequality given by Eq.(\ref{eq:positive}) is violated by an appropriate combination of different readout strategies. Specifically, the role of entanglement in quantum measurement can be verified by demonstrating that, for a specific pair of readout strategies $m_c$ and $m_r$ and any pair of eigenstates $\mid a_1 \rangle$ and $\mid a_2 \rangle$,
\begin{equation}
\label{eq:violation}
R_r(a_1,a_2) > D_c(a_1,a_2).
\end{equation}
The feasibility of this experimental confirmation of entanglement is shown by Eq.(\ref{eq:tradeoff}), which indicates that any combination of maximally coherent measurement interactions with different resolutions violate Eq.(\ref{eq:steer2}). The readout dependence in the trade-off between resolution and irreversible decoherence of a quantum measurement thus provides direct experimentally observable evidence of the generation of entanglement in the interaction stage. 
%%----

\section{Characteristics of the interaction}
\label{Sec:character}

The entangling interaction between the meter and the system is completely characterized by the conditional meter states $\mid \phi(a) \rangle$. Both the information transfer and the disturbance caused by the measurement are independent of the actual physics described by these states, and the mutual inner products of the conditional meter states are sufficient to characterize both the entanglement between the system and the meter and the availability of information in the meter system. In this section, we will consider the role of these inner products with respect to the possible readout strategies.

The resolution of the possible readouts is limited by the inner products of the conditional meter states, since the readout results must be represented by a complete basis in the Hilbert space of the meter. Therefore the inner product can be expressed in terms of the quantum statistics of the readout,
\begin{equation}
\label{eq:expand}
\langle \phi(a_2) \mid \phi(a_1) \rangle = \sum_m \langle \phi(a_2) \mid m \rangle \langle m \mid \phi(a_1) \rangle.
\end{equation}
As explained in Sec.\ref{Sec:resolve}, the maximal resolution is obtained when the inner product is equal to the Bhattacharyya coefficient, which requires that all vector components $\langle m \mid \phi(a) \rangle$ have the same complex phase. Except for an arbitrary global phase, this means that the vector components of the conditional meter state $\mid \phi(a) \rangle$ need to be real and positive, so that
\begin{equation}
\label{eq:rp}
\langle m_r \mid \phi(a) \rangle = \sqrt{P(m_r|a)}.
\end{equation}
It is easy to see that this relation can only be satisfied if all of the inner products of different conditional meter states are also real and positive,
\begin{equation}
\label{eq:positivB}
\langle \phi(a_2) \mid \phi(a_1) \rangle = |\langle \phi(a_2) \mid \phi(a_1) \rangle|.
\end{equation}
We can therefore conclude that real and positive values of all of the inner products between conditional meter states is a necessary condition for the achievement of a resolution at the limit given by Eq.(\ref{eq:RDUR}). 

In principle, there are no restrictions on the methods used to read out the meter. In particular, it is always possible to transfer the meter state without loss into a larger Hilbert space. We should therefore assume that the meter readout is not subject to any fundamental physical restrictions. This means that the readout process is completely independent of the physics of the system and the physical realization of the interaction between the system and the meter. As a consequence of this independence, the construction of an optimal readout basis $\{\mid m_r \rangle\}$ that achieves the maximal resolution of $R(a_1,a_2)=D(a_1,a_2)$ for all combinations of $a_1$ and $a_2$ is a purely mathematical problem. Specifically, we need to find a positive and real representation of the vectors $\mid \phi(a) \rangle$ that is consistent with the inner products determined by the measurement interaction. We can think of this mathematical relation as a factorization of the positive and symmetric matrix given by the inner products $\langle \phi(a_2) \mid \phi(a_1) \rangle$, as shown in Eq.(\ref{eq:expand}). To optimize the readout, we should find a non-negative real matrix $\langle m_r \mid \phi(a) \rangle$ that solves Eq.(\ref{eq:expand}) for any real and symmetric matrix $\langle \phi(a_2) \mid \phi(a_1) \rangle$. It has been shown that this problem can always be solved if a sufficiently large number of orthogonal basis states $\{\mid m_r \rangle\}$ is used \cite{Hal63,Han83}. It can therefore be proven that the positivity of the inner products of the conditional meter states given by Eq.(\ref{eq:positivB}) is a sufficient condition for the existence of an optimal readout basis with $R(a_1,a_2)=D(a_1,a_2)$ for all combinations of $a_1$ and $a_2$. 

At the opposite end of the spectrum of possibilities is a readout basis $\{\mid m_c \rangle\}$ that provides no information on the target observable, so that all $R(a_1,a_2)$ are exactly zero and the conditional output states suffer no irreversible decoherence ($D_{\mathrm{irr.}}(a_1,a_2)=0$). This condition is satisfied when the conditional probabilities in the output do not depend on $a$,
\begin{equation}
P(m_c|a_1)=P(m_c|a_2)=P(m_c). 
\end{equation}
Since the conditional meter states $\mid \phi_(a) \rangle$ are generated by different unitary transformations $\hat{U}_M(a)$ from the initial meter state $\mid \Phi_0 \rangle_M$, the condition that the distribution $P(m_c|a)=|\langle m_c \mid \phi(a) \rangle|^2$ is independent of $a$ suggests that the readout states $\mid m_c \rangle$ are eigenstates of the unitary transformation $\hat{U}_M(a)$, so that
\begin{equation}
|\langle m_c \mid \hat{U}_M(a) \mid \Phi_0 \rangle|^2 = |\langle m_c \mid \phi_0 \rangle|^2.
\end{equation}
The condition for zero irreversible decoherence can therefore be satisfied for any measurement interaction that conserves a meter observable described by a set of eigenstates $\{\mid m_c \rangle\}$. 

In general, the range of possible readouts is completely determined by the inner products of the conditional pointer states. It is relatively easy to guarantee that the readout can fully reverse the decoherence at a resolution of $R_c(a_1,a_2)=0$. The conditions that need to be satisfied by the measurement interaction to achieve full reversibility of decoherence is that the interaction Hamiltonian has product eigenstates $\mid a \rangle\otimes\mid m_c \rangle$, which means that the interaction Hamiltonian can be written as the product of two local operators, $\hat{H}=\hat{A}\otimes\hat{B}$, where $\mid a \rangle$ ($\mid b \rangle$) are the eigenstates of the operator $\hat{A}$ ($\hat{B}$) with eigenvalues of $A_a$ ($B_b$). For an effective interaction time of $t$, the conditional meter states are then given by 
\begin{equation}
\mid \phi(a) \rangle = \sum_b \exp\left(-i \frac{A_a B_b t}{\hbar}\right) \langle b \mid \Phi_0 \rangle\mid b \rangle.  
\end{equation} 
A measurement of $\hat{B}$ in the meter determines the phase changes in the off-diagonal matrix elements of the system density matrix. We can therefore recover the complete coherence by using $\{\mid m_c \rangle \}=\{\mid b \rangle\}$ as the readout basis. 

It is a bit more difficult to construct an interaction that can achieve a resolution equal to the decoherence caused by the interaction. As shown above, the necessary and sufficient condition for maximal resolution is given by Eq.(\ref{eq:positivB}), which puts a highly non-trivial constraint on the possible conditional meter states. In general, the inner products of the conditional meter states are given by a complex number,
\begin{equation}
\label{eq:Bprod}
\langle \phi(a_2)\mid \phi(a_1) \rangle = \sum_b \exp\left(-i \frac{(A_1-A_2) B_b t}{\hbar}\right) |\langle b \mid \Phi_0 \rangle|^2.  
\end{equation}
The complex phases originate from the dynamics and involve phases determined by the eigenvalues $B_b$ of the meter system. To achieve real and positive values in Eq.({eq:Bprod}), we can separate the meter system into two parts, $P1$ and $P2$, with $\hat{B}=\hat{V}_{P1}-\hat{V}_{P2}$ and $\mid \Phi_0 \rangle = \mid \Phi_{P1} \rangle \otimes \mid \Phi_{P2} \rangle$. The conditional meter states can then be given by product states of the form
\begin{equation}
\mid \phi(a) \rangle = \mid \phi_{P1}(a) \rangle \otimes \mid \phi_{P2}(a) \rangle
\end{equation} 
and the inner products of the conditional meter states are given by 
\begin{equation}
\langle \phi(a_1)\mid \phi(a_2) \rangle = \langle \phi_{P1}(a_1)\mid \phi_{P1}(a_2) \rangle
\langle \phi_{P2}(a_1)\mid \phi_{P2}(a_2) \rangle.
\end{equation}
Real and positive values can be guaranteed if the two terms on the right hand side are complex conjugates of each other,
\begin{equation}
\langle \phi_{P1}(a_1)\mid \phi_{P1}(a_2) \rangle = \langle \phi_{P2}(a_1)\mid \phi_{P2}(a_2) \rangle^*.
\end{equation}
This is automatically satisfied by the opposite sign of $\hat{V}_{P1}$ and $\hat{V}_{P2}$ in $\hat{B}$ if the two parts are otherwise identical. In terms of eigenstates $\mid v \rangle$ and eigenvalues $V_v$ for $\hat{V}_{P1}$ and $\hat{V}_{P2}$,
\begin{eqnarray}
\langle \phi_{P1}(a_1)\mid \phi_{P1}(a_2) \rangle &=& \sum_v  \exp\left(-i \frac{(A_2-A_1) V_v t}{\hbar}\right) |\langle v \mid \Phi_{P1} \rangle|^2  
\nonumber \\ 
\langle \phi_{P2}(a_1)\mid \phi_{P2}(a_2) \rangle &=& \sum_v  \exp\left(+i \frac{(A_2-A_1) V_v t}{\hbar}\right) |\langle v \mid \Phi_{P2} \rangle|^2. 
\end{eqnarray}
We therefore find that the two inner products are complex conjugates of each other for equal distributions of $v$, $|\langle v \mid \Phi_{P1} \rangle|^2=|\langle v \mid \Phi_{P2} \rangle|^2$. 

The above construction shows that measurement interactions producing conditional meter states with real and positive inner products are a realistic possibility, no matter how complicated the actual meter system gets. A measurement interaction satisfying Eq.(\ref{eq:positivB}) generates an entanglement that can be steered between a maximal readout resolution of $R(a_1,a_2)=D(a_1,a_2)$, where the limit depends only on the decoherence $D(a_1,a_2)$ caused by the measurement interaction, and a complete erasure of the measurement interaction with zero resolution associated with a full restoration of the initial coherence in the conditional output states ($D_{\mathrm{irr.}}(a_1,a_2)=0$). The initial interaction thus provides the uncertainty boundaries for the readout in the form of equal limitation for resolution and irreversible decoherence,
\begin{equation}
0 \leq R(a_1,a_2) \leq D(a_1,a_2)
\end{equation} 
and
\begin{equation}
0 \leq D_{\mathrm{irr.}}(a_1,a_2) \leq D(a_1,a_2).
\end{equation} 
A maximal violation of the steering inequality in Eq.(\ref{eq:steer2}) is obtained by comparing the maximal recovery of coherence at $D_{\mathrm{irr.}}(a_1,a_2)=0$ with the maximal resolution at $R(a_1,a_2)=D(a_1,a_2)$. In an experimental test of the relations above, the achievable difference between these two extremes may be useful figure of merit indicating the achievement of an optimal information transfer from the system to the meter in the measurement interaction. 

\section{Conclusions}
\label{Sec:conclude}

All measurement processes require a quantum mechanical interaction between the system and the meter, followed by an irreversible readout of the measurement result which only involves the meter system. The first stage of the measurement is reversible and can be fully quantum coherent, leaving the system and the meter in an entangled state whenever the input state is in a superposition of the target observable. In the analysis above, we have shown how the system-meter entanglement determines the trade-off between measurement resolution and irreversible decoherence. We obtain tight bounds on the trade-off relations by defining the measurement resolution as the Hellinger distance between the conditional probabilities associated with the eigenstates of the target observable and the decoherence as the relative reduction of the off-diagonal elements of the density matrix associated with the same pair of eigenstates. 
%%%--change 6, summary of specific results
As shown in Eq.(\ref{eq:RDUR}), the uncertainty limit of quantum measurements can then be summarized by the straightforward requirement that the decoherence caused by a quantum measurement must always be equal to or greater than the resolution achieved in the measurement. However, the actual resolution of a quantum measurement is not decided by the system-meter interaction, but depends on the specific readout strategy applied to the meter after the interaction. At the same time, the readout can also restore some information on the coherences between the eigenstates. Eq.(\ref{eq:tradeoff}) shows that, for a fully coherent interaction, the irreversible part of the decoherence is always equal to the minimum required by the resolution achieved in a specific readout.
%%%
The irreversible readout stage of the measurement can therefore be characterized as a quantum steering process, by which the choice of a particular readout strategy steers the system between maximal resolution of the target observable and maximal recovery of the coherence. 
%%%--change 6 continued
Specifically, the system-meter entanglement generated in the measurement interaction makes it possible to violate the steering inequality given by Eq.(\ref{eq:steer2}) using two different readout strategies, such that the measurement resolution for one readout exceeds the irreversible part of the decoherence for the other readout. It is therefore possible to experimentally confirm the role of entanglement in quantum measurements by an experimental observation Eq.(\ref{eq:violation}), which represent the violation of Eq.(\ref{eq:steer2}) by a specific pair of readout strategies. 
%%%

Our characterization of quantum measurements shows that the quantum mechanics of the entangling interaction determines the fundamental limits of resolution and decoherence through the non-classical correlations generated by the unitary dynamics of the interaction. Any optimized measurement interaction describes the full range of quantum steering, from complete recovery of the coherence to maximal resolution of the target observable. 
%%%--change 7, emphasis of readout non-locality
This means that the actual uncertainty trade-off between resolution and irreversible decoherence is decided by the readout dynamics inside the meter, which requires no physical contact with the system measured and could potentially happen after the system has moved to a completely different location. 
%%%
It is therefore unavoidable to consider the nature of quantum non-locality in any microscopic analysis of the trade-off between measurement resolution and disturbance. 

\section*{Acknowledgment}
This work was made possible by the Japan-India International Linkage Degree Program at Hiroshima University with support from the JSPS Inter-University Exchange Project.

\end{document}